# GNN with Model-based RL for Multi-agent Systems

Hanxiao Chen, Harbin Institute of Technology, hanxiaochen@hit.edu.cn

**Abstract.** Multi-agent systems (MAS) constitute a significant role in exploring machine intelligence and advanced applications. In order to deeply investigate complicated interactions within MAS scenarios, we originally propose "GNN for MBRL" model, which utilizes a state-spaced Graph Neural Networks with Model-based Reinforcement Learning to address specific MAS missions (e.g., Billiard-Avoidance, Autonomous Driving Cars). In detail, we firstly used GNN model to predict future states and trajectories of multiple agents, then applied the Cross-Entropy Method (CEM) optimized Model Predictive Control to assist the ego-agent planning actions and successfully accomplish certain MAS tasks.

## 1 Introduction

### 1.1 Purpose

Vision-based mechanisms have been explored well in diverse reinforcement learning applications such as playing novel Atari video games from raw pixels [1], controlling simulated autonomous vehicles with high-dimensional image observations [2] and completing robotic manipulation tasks (e.g., grasping, door-opening) based on state representations extracted from complicated vision data [3, 4]. However, it has been empirically observed that reinforcement learning from high dimensional observations such as raw pixels is sample-inefficient [5] and time-consuming. Furthermore, it's widely accepted that learning policies from physical state based features is significantly more efficient and explicit than learning from visual pixels. Thus, in this research project we focused on learning control policies from states and explored to apply a graph neural network dynamics model to predict future states in multi-agent systems, then utilize the Cross-Entropy Method optimized model-based controller to implement motion planning for the ego-agent and successfully accomplish specific MAS missions, such as multi-billiard avoidance or self-driving car scenarios (Fig. 1).

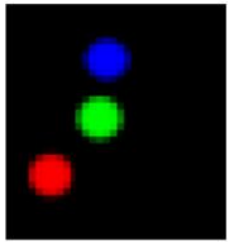 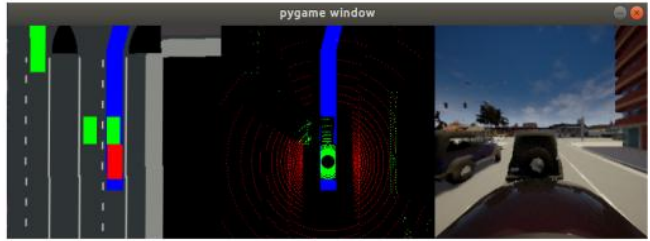

**Fig. 1.** Gym-billiard & Gym-Carla Environments.

## 1.2 Background

Inspired by [6], which presents STOVE, a creative state-space model for videos that explicitly reasons about multi-objects and their positions, velocities, and interactions, our program aims to curate another "GNN for MBRL" model based on the proposed multi-billiard simulator (Fig. 1(a)) for sample efficient model-based control in MAS tasks with heavily interacting agents. Obviously, autonomous driving is a complicated multi-agent system that requires the ego-agent to consider situations of surrounding agents and then conduct further motion planning. For this application scenario, gym-carla (Fig. 1(b)) provided by [7] can be utilized for deeper exploration. Therefore, we firstly design and validate our "GNN for MBRL" model on MAS billiard avoidance scenario which explores more possibilities of graph neural networks and model-based reinforcement learning, then try to transfer such a promising framework to real-world complicated self-driving applications.

**Graph Neural Networks.** GNN is proposed to establish representations of nodes and edges in graph data, even demonstrated notable successes in multiple applications like recommended systems, social network prediction, and natural language processing. Having observed significant potentials of GNN utilized in physics systems, we could perform GNN-based reasoning on objects and relations in a simplified but effective way by representing objects as nodes and relations as edges.

One successful GNN instance is the novel state-space model STOVE [6], which is constructed by an image reconstruction model SuPAIR [8] and a GNN dynamics model for inference, accelerating and regularizing training to push the limits of unsupervised learning of physical interactions. Thus two components are combined in a compositional manner to yield the state-space predictive model:

$$p(x,z) = p(z_0)p(x_0 \mid z_0)\prod_t p(z_t \mid z_{t-1})p(x_t \mid z_t) \qquad (1)$$

where $x$ means image observation, $z$ denotes object states. And the interface between two components are the latent positions and velocities of multiple agents. To initialize states, they model $p(z_0, z_1)$ using simple uniform and Gaussian distributions. STOVE model is trained on given video sequences $x$ by maximizing the evidence lower bound (ELBO): $\mathrm{E}_{q(z|x)}[\log p(x,z) - \log q(z \mid x)]$.

Except for basic video modeling and prediction, STOVE extends their structured and object-aware video model into reinforcement learning (RL) tasks, allowing it to be utilized for search or planning. According to their empirical evidence, an actor based on Monte-Carlo tree search (MCTS) on top of STOVE is competitive to model-free approaches such as Proximal Policy Optimization (PPO) only requiring a fraction of samples. Inspired by such RL experiments, we considered to apply the GNN model directly on states instead of high-dimensional visual data to improve the sample efficiency and predict agents' future states well, then combine it with another model-based RL method such as Model Predictive Control (MPC). Since our motivation is learning the GNN dynamics model with low-level states, in the experiment we design to train our model with ground truth states of video sequence data for multi-agent systems instead of inefficient visual data which needs to be firstly reconstructed by SuPAIR model to extract states in the original STOVE.

**Model-based Reinforcement Learning.** Model-based RL has long been viewed as a potential remedy to the prohibitive sample complexity of model-free RL. Formally, model-based RL [9] consists of two primary steps: (1) Learning a dynamics model $p(s_{t+1}|s_t,a)$ which can generate future states based on current states and actions. (2) Conducting the planning algorithm to learn a global policy or value function and act the environment well. STOVE utilized Monte-Carlo tree search (MCTS) to obtain a policy based on the world model which is leveraged as a simulator for planning. And they discovered MCTS combined STOVE could exactly outperform the model-free PPO algorithm in the multi-billiards avoidance task.

## 2 Method

### 2.1 Framework

Fig. 2 intuitively explains two important stages in our "GNN for MBRL" method: (1) **GNN dynamics model training stage**, which will be input the offline recorded video sequences data or low-dimensional states and trained for video prediction; (2) **Motion planning stage** with CEM-based Model Predictive Control (MPC), that we utilize a feed-back control algorithm integrated with a Cross-Entropy Method optimizer to interact with the billiard environment and plan reasonable actions to control the ego-agent with purpose of avoiding collisions between other agents.

Also, we have marked two different cases in the GNN dynamics training stage, one is the "Action-conditioned case" belonging to the original STOVE model-based control approach that trains GNN align with an object reconstruction SuPAIR model on visual observation data, another one is "Supervised RL case" that we primarily proposed to address RL tasks directly on low-level states without the processed visual observation information. But the framework for two cases is firstly training GNN dynamics models for multi-agent future states prediction, then integrates the trained GNN model into the following model-based RL section to control the ego agent for correct motion planning (e.g., avoiding billiard collisions, planning traffic trajectories).

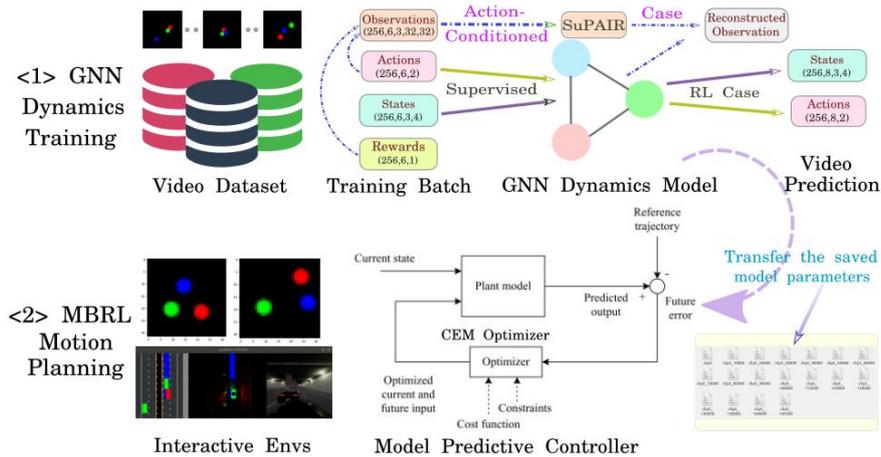

**Fig. 2.** GNN for MBRL whole pipeline.

Transferred to the second part, MPC is a feedback control algorithm that applies a model to make predictions about future outputs of a certain process. It possesses several advantages like handling multi-input multi-output (MIMO) systems with more constraints and it can easily incorporate future reference information into the control problem to improve controller's performance. Therefore, we established a continuous version of multi-billiard environment based on original discrete STOVE scenarios for data collection. And it's possible for us to combine the previously trained GNN model with MPC and investigate if this method can successfully address MAS tasks.

### 2.2 Data Generation

STOVE [6] proposed a structured object-aware physics prediction model based on the heavily interacting billiards simulation environments with diverse modes including *avoidance, billiards, gravity, and multi-billiards* (Fig. 3). However, their environment script is just attached to the whole Github repository. Interested by such amazing multi-agent systems, firstly we wrapped them into the gym environment style "gym-billiard" (Fig. 1(a)) which can be easily imported by Python API, even assist other researchers to understand this physical system and design efficient algorithms.

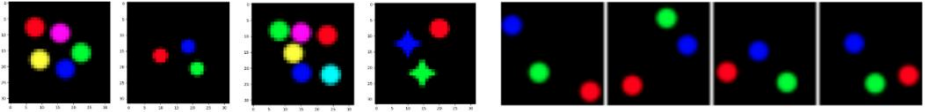

**Fig. 3.** Multiple scenarios in the gym-billiard environment (left); Avoidance task (right).

To emphasize, our project especially focuses on the interesting avoidance billiard scenario, which means the red ball serves as the ego-object and the RL agent needs to control it to avoid collisions between other balls. In the original STOVE paper, the ego-ball is controlled by nine actions, which correspond to moving in one of the eight (inter)cardinal directions and staying at rest. Also, a negative reward of -1 is given whenever the red ball collides with one of the others. Since STOVE trains exactly on video data, so firstly we obtained the avoidance sequences datasets by applying "generate_billiards_w_actions" function in the envs.py. Separately, we generated 1000 sequences of length 100 for training (avoidance_train.pkl) and 300 sequences of length 100 for testing (avoidance_test.pkl) with a random action selection policy. It is also possible to apply the Monte Carlo action policy to generate video sequence data. Furthermore, the pixel resolution of such dataset was 32*32 and we set the mass of the ball to be 2.0 with 1.0 radius size in the environment config file.

However, we discover that most model-based RL environments often provide continuous actions for agents with the action space 2 rather than 9 discrete choices. Thus, we are motivated to make changes in envs.py to get continuous action values, where the red ball can be controlled by (1,2) dimensional numpy values ranged in (-2,2). In this case, each value in actions means the acceleration of the ball in x and y directions. Similar to the discrete mode, we produced continuous datasets with the random action policy following the uniform distribution within the range of (-2,2). Furthermore, we even obtained the avoidance gif and found it performed an excellent video for three interactive balls. Table 1 shows basic information (e.g., action_space, average rewards) of two different datasets. As we can see, average rewards of training and testing data in the continuous mode are smaller than that of the discrete condition, which means three balls have more collisions and they heavily interact with each other in the continuous environment. More detailedly, each pickle file restores image observations, actions, states, dones, rewards for training and testing sequence data in the dictionary form, which will be applied in the following phase.

**Table 1.** Basic comparisons of the continuous and discrete datasets.

| Data Mode | Action space | Actions dtype | Average Rewards of training data | Average Rewards of testing data |
|---|---|---|---|---|
| **Discrete** | 9 | One-hot mode | -17.276 | -16.383 |
| **Continuous** | 2 | Numpy array | -18.93 | -18.71 |

### 2.3 GNN Dynamics Model Training

As demonstrated in our training pipeline (Fig. 2), we intended to utilize a supervised learning method which performs training on the ground-truth states rather than high-dimensional image data to improve the sample efficiency, then combines the saved GNN model with CEM optimized MPC to predict future states of billiards in the interactive Avoidance environment and replan ego-agent's actions. Then we will train two different cases on both Discrete and Continuous Avoidance datasets: (1) Action-conditioned case, where [6] has extended STOVE to reinforcement learning (RL) settings with two changes including a conditional prediction based on state & action, and a reward prediction to yield the distribution $p(z_t, r_t | z_{t-1}, a_{t-1})$. (2) Supervised RL case, where we consider the real states (batch_size, frame_number, 3, 4) just including object positions and velocities as the input of GNN dynamics model for physics predictions by replacing the SuPAIR-inferred states. Thus, the model can directly learn to predict future states of multiple agents instead of firstly extracting agent state representation from visual data with SuPAIR model.

### 2.4 GNN with MBRL

After training and saving GNN dynamics model, we design to follow the traditional Model-based RL pipeline to combine the trained GNN model with CEM-optimized Model Predictive Control (MPC) approach [9] to investigate its performance on the interactive continuous gym-billiard avoidance task. Thus, the saved GNN model is integrated into the Model-based RL algorithm loop to predict future states of multiple agents from the learned real transitions in the MAS environment, then MPC searches for an optimal action sequence for our ego-agent within the MAS scenario under the learned model and executes the first action of that sequence, discarding the remaining actions. Typically this search is repeated after every step in the gym-billiard scenario, to account for any prediction errors and rewards by the model and to get feedback from the interactive environment.

For further analysis and validation for our proposed "GNN with MBRL" method, we firstly conducted experiments on discrete ball datasets with MCTS as in STOVE [6] and saved some empirical videos. Secondly, for the continuous case, we creatively implemented the trained GNN dynamics model with the CEM optimized MPC and compare the performances with random and ground_truth situations.

## 3 Results

### 3.1 GNN Training Results

While collecting discrete and continuous train or test sequence data via our revised gym-billiard API environment, we save them in pickle files which restore image observations, actions, states, dones, rewards for multi-billiards in the dictionary form. After that, we will train GNN dynamics models in two different cases (1) Action-conditioned case, which follows [6] to train GNN on given video sequences with a visual reconstruction model SuPAIR. (2) Supervised RL case, where we consider real states (batch_size, frame_number, 3, 4) just including object positions and velocities for multi-agents as the input of GNN dynamics model for physics predictions by replacing the SuPAIR-inferred states. And we conduct training in two conditions for 500 epochs and restore important model parameter files.

Certainly training time for the Supervised condition (~8 hours) is far less than the Action-conditioned case (~25 hours). But the training processes for Discrete and Continuous situations did not show too much difference. And one of our important discoveries is that the novel GNN model can work exactly well on the new action space 2 as for 9 discrete actions, even with no changes in the original GNN network architecture. Thus, this GNN dynamics framework can perfectly achieve the unity training for both Discrete and Continuous billiard avoidance environments.

Intuitively, Fig. 4 is the screenshot for the folder that we obtained after training two cases, where "checkpoints" saved the model parameters, "gifs" contained the real, generated reconstructed or rollout videos, "states" included some crucial files of states or rewards after certain training timesteps. In addition, the generated data can refer to: https://pan.baidu.com/s/1evwHqrtVJE5EM46wJHxaZg (key: hru9).

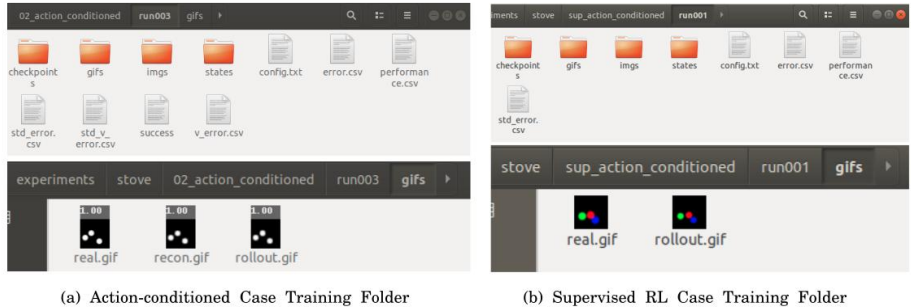

(a) Action-conditioned Case Training Folder  (b) Supervised RL Case Training Folder

**Fig. 4.** Screenshot for training results of two cases.

Fig. 5 represents what we've obtained in the performance.csv of two different cases after training. Obviously, the Action-conditioned training situation has restored more numerical metrics like elbo, reward, min_ll, v_error than the Supervised RL csv file. Another interesting thing for both cases is that they save such logging files in a joint training and testing style, which means the training and testing metrics for GNN will appear repeatedly. From each individual Action-conditioned file, you can find each row corresponds to different types, where "z", "z_sup", "z_dyn" belongs to the SuPAIR image model's training results, but the "z_roll", "z_sup_roll", "z_dyn_roll" denote testing consequences for the GNN rollout video sequences. Selected from such metrics, we focused on 4 different values: reward, elbo, error, and v_error, since here the reward means the MSE loss between the GNN predicted rewards and ground truth rewards instead of the traditional award definition. So the lower the reward is, the model will be trained better. Elbo is very important because the GNN is trained on given video sequences x by maximizing the evidence lower bound (ELBO). Error and v_error separately denote the position and velocity MSE loss between the predicted and true values.

**Fig. 5.** Performance.csv partial-shown files for 4 different training cases.

**Action-conditioned case.** GNN training results for Action-conditioned condition are presented in Fig. 6 and here we show the "z" type of data to explain each metric since it contains the full information of "z_sup" and "z_dyn". Obviously, we discover the reward MSE loss is decreasing in both continuous and discrete conditions, but the continuous reward error declined from 0.48 to 0, whereas the discrete from 0.16 to 0. As our GNN model training motivation is maximizing ELBO and the training curves indeed increase this metric significantly from 450 to 3600.

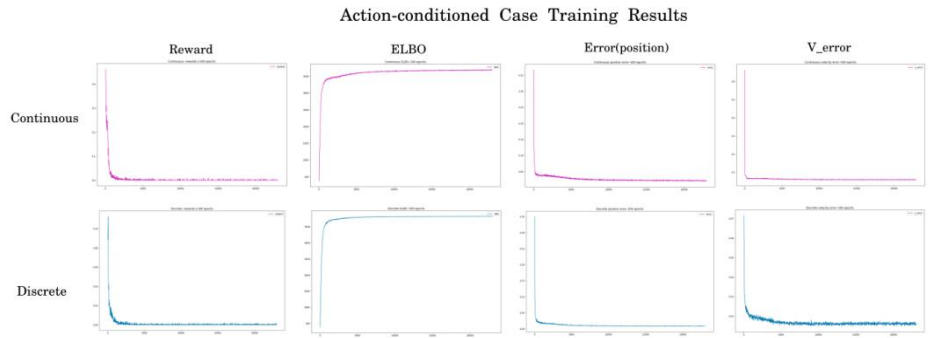

**Fig. 6.** Training curves for Action-conditioned case.

In addition, we pay attention to the position prediction error (Error) and velocity error (V_error) since these two variables determine the actual states of MAS systems. Surely they both decrease while training but we also find the continuous position error is close to the discrete one, but the velocity error shows a large amount of difference, where the continuous V_error drops from 0.65 to 0.05 but the discrete one decreases from 0.07 to 0.01. This seems a little strange but it's related to our new action space definition: two values of (1, 2) numpy array range in (-2, 2) separately means the acceleration in x & y directions, so it may causes much gap or disparity in the velocity prediction. According to the generated real, recon, and rollout gifs, we discovered that they are reasonable and the billiards in rollout videos run much slowly than others. Generally speaking, such four metrics training curves exactly meet the criteria of a reasonable GNN dynamics model for the following RL task.

**Supervised RL case.** As shown in Fig. 2 pipeline, we directly input the ground truth states and actions to teach the GNN dynamics model predicting future possible states well. Similarly, the performance.csv records the total, prediction, and reconstruction errors of the system iteratively for training and testing. As the Supervised RL case (Fig. 7) does not require any image reconstruction operations on true states, the Reconstruction_error always equals to 0. "Prediction_error" is another important metric to evaluate the model prediction ability. Compared with the continuous case, the discrete one performs much better since it drops the loss from 1.1 to 0.1. For the "Total_error" which is connected to multiple factors, we find the continuous loss keeps in a stable range but the discrete case firstly shows an apparent downtrend and then keeps stable. To further validate the training results, we checked the generated real and rollout gifs. Obviously, all the rollout videos make the balls run much slowly but the ego-red ball exactly reflects reasonable avoidance ability. Therefore, based on comprehensive analysis, we considered that the trained Supervised RL model can be utilized for the following model-based reinforcement learning control phase.

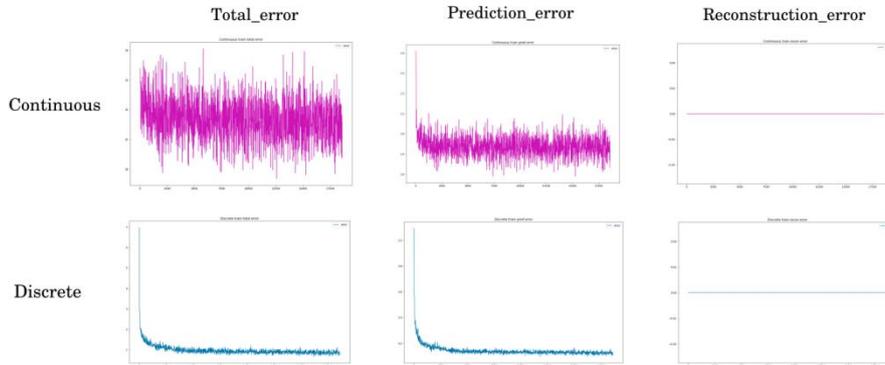

**Fig. 7.** Training curves for Supervised RL case.

### 3.2 GNN with MBRL

In this section, we provide experimental results of the integration of GNN dynamics model and Model-based control methods on the gym-billiard avoidance MAS task. At first, we follow the "Model-based Control" framework in STOVE [6] on discrete ball datasets with MCTS and obtained qualitative videos for visualization. Secondly, for the continuous datasets, we creatively integrated our trained GNN dynamics model into the CEM optimized MPC method and compare their performances with random and ground_truth situations.

**STOVE with MCTS (Discrete).** We successfully re-implemented the combination of STOVE with MCTS on our generated discrete datasets. Also, we've changed the mass of multiple ball agents from 1.0 to 2.0 to produce two different pickle files and respectively trained the GNN dynamics models as the proposed Action-conditioned training cases. In detail, while running MCTS, we used the GNN model to predict future sequences for 100 paralleled interactive gym-billiard avoidance environments with the running length of 100, and the maximal rollout depth for each tree search equals to 10. For this experiment, we focused on the mean collision rate of 100 environments and Fig. 8 clearly shows the calculated collision rate of each single scenario. Also, we saved 100 interesting gifs after running MCTS to help us observe how the red ego-ball interact with other agents. Surely, this red ball performs much better to avoid collisions with MCTS and achieves lower mean collision rate for paralleled environments.

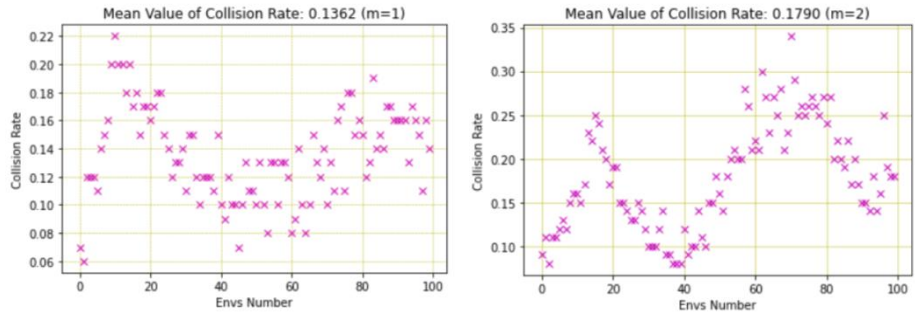

**Fig. 8.** STOVE with MCTS reward distribution.

**GNN with CEM based MPC (Continuous).** To achieve our proposed "GNN with MPC" approach, we load the saved supervised-RL case GNN ckpt files and apply the model parameters for prediction. Another challenging problem for implementation is appeared on data structure for the connection of trained model and the MPC interface, we must ensure that the GNN model can exactly conduct predictions based on the current states without bugs. Surprisingly, we addressed this issue well by checking the code row to row, changing the cuda device config to cpu and many variables' dtype in both supairvised/dynamics.py and video_prediction/dynamics.py.

After solving above tough issues, we started to integrate the GNN trained models into the classical CEM optimized MPC, and conducted experiments on two different billiard environments with different epochs and horizons, even compute the "reward" with other two baselines (random + ground_truth). Here, "reward" means the average collisions happening in each epoch with different horizons since we make some changes for the envs.py in mpc folder so that a positive reward of 1 is given whenever the red ball collides with other balls. Table 2 collected different reward (similar to collision rate) values for our GNN_MPC model with two training baselines (random & ground_truth). To emphasize, the "random" case means that the future actions will be generated randomly instead of using the MPC controller chosen actions to conduct motion planning. Whereas, the "ground_truth" denotes applying the true interaction billiard environment to generate next states rather than via the GNN model. Another important thing is that the "m=1" version task differs a little from that "m=2" since the "m=1" GNN model is initially trained on the old continuous datasets which generate the red ball's movement not as flexible as that in the new "m=2" training data. Thus we call it the "Old_env" to contrast with the "New_env". According to Table 2, we discover that collision rates in the "GNN_MPC" case are much lower than "Random", even much close to the "ground_truth" condition.

**Table 2.** Reward (Collision Rate) for two envs with different baselines.

| Envs | Epochs | Horizons | GNN_MPC | Random | Ground_truth |
|---|---|---|---|---|---|
| m=1 | 100 | 50 | **0.0558+0.0012** | 0.2790+0.025 | 0.0707+0.066 |
| m=1 | 50 | 100 | **0.0565+0.0008** | 0.3543+0.0445 | 0.0408+0.0392 |
| m=2 | 100 | 50 | **0.0648+0.001** | 0.2420+0.0178 | 0.0505+0.0480 |
| m=2 | 50 | 100 | **0.0455+0.0008** | 0.2690+0.0350 | 0.0612+0.0575 |

In addition, we've recorded each epoch's "reward" values in the csv files for 4 cases mentioned in Table 2 and plot them together in Fig. 9. As we can see, the plotting results correspond to our calculated collision rewards and 1.0 means that the ego-object collides with other balls in all horizons of the whole epoch. So if the value of average returns is much lower, the method's performance is much better. From this perspective, we surprisingly find our new proposed "GNN_MPC" model ("STOVE") exceeds the "Random" case significantly, which means the MPC controller performs really well to choose a high-quality action. Since the results of "GNN_MPC" are extremely close to the "Ground_truth" case, it is further indicated and validated that our trained GNN dynamics model woks exactly well to predict future states of multi-object systems as the same to the ground_truth interactive environment.

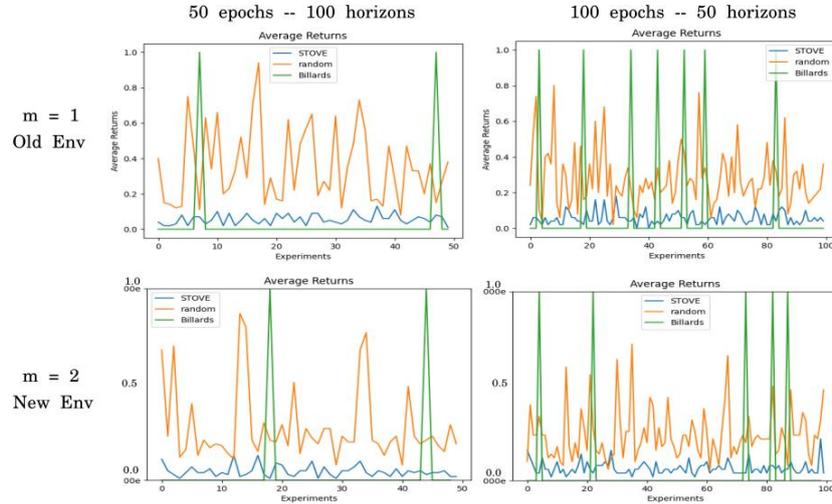

**Fig. 9.** GNN with MPC experimental results.

## 4  Conclusions

In summary, we proposed "GNN for MBRL" idea which combines the novel graph neural network (GNN) dynamics model with CEM optimized Model Predictive Control (MPC) on the gym-billiard avoidance MAS task. Furthermore, we not only conducted extensive experiments on "Action-conditioned" case as in STOVE with MCTS via original discrete datasets, but also explored and evaluated "Supervised RL" GNN dynamics model with CEM optimized MPC on our revised continuous datasets. According to empirical results, we discovered that our creative idea performs well to predict future video sequences and control the ego-agent to exactly address certain RL tasks, thus this model may be applied and extended on much complicated multi-agent systems such as the gym-carla autonomous driving environment.